\documentclass[paper]{elsart}
\journal{Physica D}
\usepackage{natbib}\usepackage{graphicx}\usepackage{amsmath}\usepackage{amssymb}
\begin{document}\begin{frontmatter}\title{Standing Waves in a Non-linear 1D Lattice : Floquet Multipliers, Krein Signatures, and Stability}
\author[sp]{Subhendu Panda}\ead{subhendup@vsnl.net}
\author[anl]{Anindita Lahiri}\ead{l\_ani1@rediffmail.com}
\author[tkr]{Tarun K. Roy}\ead{tarun@theory.saha.ernet.in}
\author[al]{Avijit Lahiri\corauthref{cor}},\corauth[cor]{Corresponding author.}\ead{a\_l@vsnl.com}
\address[sp]{Dept of Physics, Vidyasagar Evening College, Kolkata 700 006, India}
\address[anl]{Dept of Physics, Behala College, Kolkata 700 060, India; Institute of Interdisciplinary Research, Kolkata 700 005, India}
\address[tkr]{Saha Institute of Nuclear Physics, 1/AF, Bidhannagar, Kolkata 700 064,
 INDIA}
\address[al]{ Dept of Physics, Vidyasagar Evening College, Kolkata 700 006, INDIA}
\date{}
\begin{abstract}
We construct a class of exact commensurate and incommensurate standing wave (SW) solutions in a piecewise smooth analogue of the discrete non-linear Schr\"{o}dinger (DNLS) model and present their linear stability analysis. In the case of the commensurate SW solutions the analysis reduces to the eigenvalue problem of a transfer matrix depending parametrically on the eigenfrequency. The spectrum of eigenfrequencies and the corresponding eigenmodes can thereby be determined exactly. The spatial periodicity of a commensurate SW implies that the eigenmodes are of the Bloch form, characterised by an even number of Floquet multipliers. The spectrum is made up of bands that, in general, include a number of transition points corresponding to changes in the disposition of the Floquet multipliers. The latter characterise the different band segments. An alternative characterisation of the segments is in terms of the Krein signatures associated with the eigenfrequencies. When one or more parameters characterising the SW solution is made to vary, one occasionally encounters collisions between the band-edges or the intra-band transition points and, depending on the the Krein signatures of the colliding bands or segments, the spectrum may stretch out in the
complex plane, leading to the onset of instability. We elucidate the correlation between the disposition of Floquet multipliers and the Krein signatures, presenting two specific examples where the SW possesses a definite window of stability, as distinct from the SW's obtained close to the anticontinuous and linear limits of the DNLS model.
\end{abstract}
\end{frontmatter}
\section{Introduction}
The question of enumerating and classifying the possible excitations in a 
discrete non-linear lattice (e.g., a Fermi-Pasta-Ulam(FPU) chain) is too complex to be even 
addressed in its entirety. However, important classes of excitations have 
already been identified and studied exhaustively, and their physical 
implications are in the process of being worked out. One such class consists 
of localised solutions, namely the discrete breathers with periodic or 
quasi-periodic time-variation, while a related class includes the travelling
breather solutions. A third class of solutions can be described as non-linear
spatially periodic travelling waves analogous to the phonon modes in linear 
lattices. The existence of such travelling waves has been established in non-linear discrete Klein-Gordon (NDKG)
lattices as well as in FPU chains~\cite{ref1,ref2}  close to the linear limit (i.e., for
small amplitudes) as also in the vicinity of the so called anti-continuous 
limit (small inter-site coupling). In the latter case, the travelling waves 
were constructed as an infinite array of breathers (multi-breathers) with
an appropriate phase introduced at each breather site.
\vskip .5cm
\noindent
More recently, commensurate and incommensurate {\it standing wave} (SW) solutions have
also been investigated in the literature, once again close to the two limits 
referred to above~\cite{ref3,ref4} . Close to the linear limit, small amplitude SW 
solutions in the NDKG chain have been obtained from the DNLS approximation 
by referring to  orbits in a small neighbourhood of a fixed point in a related 
area-preserving mapping. In the weak-coupling situation, on the other hand,
one obtains the SW solutions by continuation from the anti-continuous limit 
of a solution with an appropriate coding sequence. In either of these 
approaches one can obtain SW solutions both commensurate and incommensurate with
the underlying lattice. Even chaotic spatial structures are not ruled out.  
\vskip .5cm
\noindent
The question of stability of these standing wave solutions is a more complex 
one. In the case of the travelling waves, it appears that they are afflicted 
with a type of modulational instability, as a result of which they break up
into an array of localised pulses. The SW's in general also happen to be 
linearly unstable close to the small amplitude limit. For the commensurate
SW's the eigenfrequencies form bands on the real axis, but these bands have
a tendency to overlap and extend into the complex plane leading to oscillatory
instabilities for arbitrarily small amplitudes. The spectra for incommensurate
SW's, on the other hand, are generaly cantor-like and require a more 
sophisticated analysis~\cite{ref3,ref4}, but here again one finds that the SW's
with even arbitrarily small amplitudes are prone to oscillatory instability.
\vskip .5cm
\noindent
In the present paper we present an exact construction of commensurate and 
incommensurate SW's in a piece-wise smooth (PWS) system resembling 
the DNLS
model and perform an exact linear stability analysis for the commensurate 
SW's. The linearised evolution equation for an SW in the model depends 
essentially on two parameters - the breather frequency $\omega$ and
a nonlinearity parameter $\gamma$ and, in contrast to breathers obtained 
perturbatively from the linear limit, one can identify well-defined regions in 
the $ \omega$-$\gamma $ parameter space where the SW's are linearly stable. 
In other words, away from the linear and anti-continuous limits there do 
exist stable commensurate SW solutions - the main result of this paper. 
In addition, the bands of eigenfrequencies arising from the linearised 
evolution equation which we calculate in arriving at the stability result 
present interesting features. While in general there exist gaps between the 
edges of successive bands and the width of the band gaps varies with the 
parameters $ \omega,~ \gamma$, instability {\it does not necessarily occur} through 
a collision of the band-edges (i.e., shortening of a band-gap to zero width). 
As we shall see, there may exist, in addition to the band-edges, a number of 
transition points (see below) {\it within} the bands and collisions of these 
intra-band transition points may equally well account for the onset of 
instabilty. The 
model under consideration being explicitly solvable, the detailed structure of 
a band can be worked out, including the location of the band-edges and 
transition points as also the eigenmodes associated with the different band 
segments. Additionally, the different band segments can be characterised in terms of the Krein signatures~\cite{ref5,ref6,ref7} associated with the eigenfrequencies, and the correlation between the various possible dispositions of Floquet multipliers (see below) and the Krein signatures can be looked at. This gives us insight into infinite dimensional linear Hamiltonian 
systems with Bloch type eigenfunctions in the context of the theory of Hamiltonian Hopf bifurcation~\cite{ref8,ref9,ref10} in  infinite dimensional systems (see also~\cite{ref7}).
\vskip .5cm
\noindent
In section 2 below, we present the piecewise smooth DNLS model and briefly
recall results from a couple of earlier papers~\cite{LPR1,LPR2} relating to the single-site and
multi-site breather solutions. Section 3 presents the exact commensurate and
incommensurate SW solutions, obtained by extension of the multi-site 
breathers where the breathers form an infinitely extended periodic or quasiperiodic array. Section 4 is devoted to the formulation of the stability problem with special reference to the commensurate SW solutions where the problem of 
calculating the spectrum of eigenfrequencies is reduced to the consideration 
of the eigenvalue problem of a $4\times 4$ complex {\it transfer matrix} depending 
parametrically on the eigenfrequency $p$ of the original problem as also on
the breather parameters $\lambda$ (related to the breather frequency $\omega$),
$\gamma$. We indicate how the bands, together with their edges and transition 
points, are determined by the parametric bifurcation of the  transfer 
matrix. This section also elucidates the correlation between the Floquet multipliers and the Krein signatures associated with the different band segments in the context of the linearised variational equations which constitute an infinite dimensional linear Hamiltonian system, and their role in characterising the onset of instability.
In section 5, we specialise to a couple of particular cases involving SW's 
with specified spatial periodicity and phase characteristics. In both these cases 
the onset of instability is associated with a collision among intra-band 
transition points, and we obtain explicitly the window of stability in the 
$\lambda$-$\gamma$ plane. This section also contains concluding remarks.
\section{Breather solutions in a DNLS-like model}
In~\cite{LPR1,LPR2} we considered a piecewise smooth DNLS-like model given by 
\begin{subequations}
\begin{equation}
i \frac{d\psi_{n}}{dt}+(\psi_{n+1}+\psi_{n-1})+f(|\psi_{n}|)\psi_{n}=0,
\label{eq:onea}
\end{equation}
\noindent with 
\begin{equation}
f(x)=\gamma (1-\frac{1}{x}) \ominus (x-1), \label{eq:oneb}
\end{equation}
\end{subequations}
\noindent where $\ominus$ stands for the Heavyside step function and $\gamma$ is a 
parameter characterising the strength of non-linearity. The non-linear term also
involves a threshold parameter which we have scaled to unity. With reference to this threshold parameter we characterise lattice sites as being either 
{\it high} or {\it low} depending on whether $|\psi_{n}|>1$ or $|\psi_{n}|<1$ 
respectively.
As indicated in~\cite{LPR1,LPR2} the model is qualitatively similar to the cubic DNLS model (though the latter does not explicitly involve a threshold 
parameter) and possesses breather solutions of similar features. In particular,
we presented explicit construction of single site and two-site breathers as also
an explicit stability analysis of these breathers. 
\vskip .5cm
\noindent
The single site breather solution is characterised by two parameters : a 
spatial decay rate $\lambda$ (related to the breather frequency as 
$\omega=-\lambda-\frac{1}{\lambda}$) and the nonlinearity parameter $\gamma$,
and is given by 
\begin{subequations}
\begin{equation}
\psi_{n}={\bar\phi}_{n} e^{-i\omega t}, \label{eq:twoa}
\end{equation}
\begin{equation}
{\bar\phi}_{n}= \frac{\gamma}{\gamma+\lambda-\frac{1}{\lambda}} \lambda^{|n|}. \label{eq:twob}
\end{equation}
\end{subequations}
The parameter $\gamma$ is to be larger than a certain threshold value 
(depending on $\lambda$) for this breather solution to exist.
The linearisation of \eqref{eq:onea}, \eqref{eq:oneb} around \eqref{eq:twoa}, \eqref{eq:twob} in the rotating frame was then shown to 
possess a spectrum consisting of a single band with extended eigenmodes 
and an isolated point corresponding to a localised mode (in addition to a trivial 
phase mode with eigenfrequency zero). As the parameters $\lambda$ and $\gamma$
are made to vary, the isolated point in the spectrum gets shifted while the 
band remains fixed in the frequency scale and, as a certain stability border in 
the $\lambda$-$\gamma$ plane is crossed, the isolated point in the spectrum 
approaches zero (the `double-zero' configuration) with a consequent onset of
stationary instability.  Interestingly, there exists a certain critical value of the spatial
decay rate $(\lambda_c)$ above which the breather becomes intrinsically 
unstable. 
\vskip .5cm
\indent
In~\cite{LPR2} the model was also shown to possess $2$-site monochromatic breather 
solutions of the form (\ref{eq:twoa}) with {\it high} sites located at, say, 
$n=0$ and $n=N$, 
the amplitude at the {\it high} sites being 
\begin{equation}
b=\frac{\gamma (1-\lambda^{2N})}{\gamma (1-\lambda^{2N})-(\frac{1}{\lambda}-\lambda)(1-\lambda^N e^{i \delta})} \label{eq:three}.
\end{equation}
Here $\delta$ is the phase difference between the two high sites, which can be
either $0$ or $\pi$, and once again $\gamma$ is to be larger than a certain 
threshold value depending on $\lambda$ for the breather to exist. An exact
linear stability analysis was performed for the two-site breathers. The 
in-phase breather $(\delta=0)$ was found to be intrinsically unstable while,
for anti-phase breather ($\delta=\pi$), one obtains a band of extended modes
together with a pair of localised {\it symmetric} modes and another pair of localised
{\it antisymmetric} modes for parameters $(\lambda,\gamma)$ away from a stability 
border (for given $N$). As the parameters are made to vary, the isolated eigenfrequencies
get shifted and, at the stability border the breather is destabilised through
a Krein collision involving the symmetric eigenmodes. 
\vskip .5cm
\noindent
The model (\ref{eq:onea}),(\ref{eq:oneb}) admits of many other breather solutions 
including, possibly, 
spatially random ones. We present below a class of exact commensurate and 
incommensurate standing wave (SW)
solutions in the model and then perform their linear stability analysis, with 
special reference to the commensurate SW's. 
 
\section{Commensurate and Incommensurate SW solutions}
\vskip .5cm
\noindent
We consider a class of breather solutions for which the {\it high} sites are 
located in a periodic array at $ n=0,\pm N, \pm 2N, \pm 3N, \ldots$.
However, successive high sites are allowed a phase difference $\delta$ that may
or may not be a rational multiple of 2$\pi$ ({\it cf.} the $2$-site breather solution 
where $\delta$ is either 0 or $\pi$). In between the {\it high} sites there 
occurs a succession of {\it low} sites where one has 
\begin{equation}
i \frac{d\psi_{n}}{dt}+(\psi_{n+1}+\psi_{n-1})=0. \label{eq:four}
\end{equation}  
Considering monochromatic breathers of the form \eqref{eq:twoa}, \eqref{eq:four} gives
\begin{equation}
{\bar\phi}_{n}=\alpha \lambda^n + \beta \lambda^{-n} \label{eq:five}, 
\end {equation}
where $\lambda$ satisfies
\begin{equation} 
\lambda+\frac{1}{\lambda}=-\omega \label{eq:six},
\end{equation}
and where we choose without loss of generality $|\lambda|<1$. Note that 
$|\omega|>2$, i.e., the breather frequency lies outside the linear phonon 
band. In the following we restrict ourselves to $\lambda>0$ ($\omega<-2$)
for the sake of concreteness (corresponding results are obtained for 
$\lambda<0$). 
\vskip .5cm
\noindent
To be more specific, we look for monochromatic SW solutions of the form (\ref{eq:twoa}) 
where {\it high} sites are 
located in a periodic array on the lattice at, say, $n=0,\pm N, \pm 2N,\ldots
(N\geq 2) $. Each high site is flanked on either side by a {\it low} site and the 
magnitude $|\psi_{n}|$ at each high site is the same, say, $b$. However, we 
allow for a phase difference $\delta$ between successive high sites. We note
from \eqref{eq:onea}, \eqref{eq:oneb} that for a low site (i.e., one with $|\psi_{n}|<1$),
\begin{equation}
{\bar\phi}_{n}+{\bar\phi}_{n+1}+{\bar\phi}_{n-1}=0 \label{eq:seven},
\end{equation}
i.e., ${\bar\phi}_{n}$ is of the form \eqref{eq:five},
where $\lambda$ satisfies \eqref{eq:six} and $\alpha$, $\beta$ are appropriate constants. For a
{\it high} site with $|\psi_{n}|=b>1$, on the other hand, 
\begin{equation}
\omega b e^{i \theta_{n}} + {\bar\phi}_{n+1}+{\bar\phi}_{n-1}=-\gamma b e^{i \theta_{n}} (1-\frac{1}{b}) \label{eq:nine},
\end{equation}
where $\theta_{n}$ denotes the phase at the site under consideration. Taking
together all these requirements, one arrives at the following SW solution:
\begin{equation}
{\bar\phi}_{kN+m}=b e^{i k \delta} \left(\frac{1-e^{i \delta} \lambda^N}{1-\lambda^{2N}} \lambda^m + \frac{e^{i \delta}-\lambda^N}{1-\lambda^{2N}} \lambda^{N-m}\right) \label{eq:ten},
\end{equation}
where the index $n$ for a site has been expressed in the form 
\begin{equation} 
n=kN+m ~~(0\leq m<N,~~ k=0, \pm 1, \pm2,\ldots),\label{eq:eleven}
\end{equation}
with
$k, m$ integers determined
uniquely by $n$ (for given $N$). Note that, while the phase advance $(\delta)$
is uniform between successive {\it high} sites $(n=kN, n=(k+1)N)$, it is not 
necessarily so for arbitrarily chosen pairs of successive
sites $n,n+1$. 
\vskip .5cm
\noindent
It is also important to note that the standing wave solution 
(\ref{eq:twoa}), (\ref{eq:ten}) is valid only if $\gamma,~\omega$ satisfy appropriate constraints 
corresponding to the consistency conditions
\begin{equation}
|{\bar\phi}_0|>1,~~~~~|{\bar\phi}_1|<1, \label{eq:twelve}
\end{equation}
which imply, in general, that $\gamma$ is to be larger than a certain threshold 
value $\gamma_1(\lambda)$ depending on $N$ and $\delta$ (as already mentioned, similar conditions 
also apply for single-site and two-site breathers obtained in~\cite{LPR1,LPR2}). 
Expressions of $\gamma_1(\lambda)$ for the particular cases $N=2,~\delta=\pi$, and 
 also for $N=3,~\delta=\pi$, will be found below. 
\vskip .5cm
\noindent
Evidently, (\ref{eq:ten}) represents a spatially periodic, or 
{\it commensurate} standing wave solution if $\delta$ is a rational  multiple 
of $2\pi$, say, $ \delta=(\frac{r}{q})2\pi$ ($r,~q$, co-prime integers, $q>0$), 
in which case both
the amplitude and phase are repeated at intervals of $\tilde{N}=q N$ lattice sites.
On the other hand, if $\frac{\delta}{2\pi}$ is irrational, (\ref{eq:ten}) 
represents an {\it incommensurate} standing wave since the phase of $\psi_n$ 
never repeats itself, but comes arbitrarily close to any preassigned value 
$\bar{\delta}$ for some $n$ or other. 
\vskip .5cm
\noindent 
In section (5), we consider for the sake of concreteness two particular 
{\it commensurate} solutions with $\delta=\pi$ and with $N=2,~3$ respectively. 
The solution for $N=2,~\delta=\pi~(\tilde{N}=4)$ reads (up to translation by a 
lattice site)
\begin{subequations} 
\begin{eqnarray}
{\bar\phi}_0=-{\bar\phi}_2=b\label{eq:thirteena},\\
{\bar\phi}_1={\bar\phi}_3=0 \label{eq:thirteenb},\\
{\bar\phi}_{n+\tilde{N}}={\bar\phi}_n, \label{eq:thirteenc}
\end{eqnarray}
where 
\begin{equation}
b=\frac{\gamma}{\gamma+\omega}, \label{eq:thirteend} 
\end{equation}
and where $\gamma$ is to satisfy the consistency condition 
\begin{equation}
\gamma>\gamma_1(\lambda)|_{N=2,~\delta=\pi} =-\omega . \label{eq:thirteene}
\end{equation}
\end{subequations}
\vskip .5cm
\noindent
Recall that in this paper we restrict to $\omega<-2$, while similar results are 
obtained for $\omega>2$.
In a similar manner, one has, for $N=3$, $\delta=\pi$ (periodicity 
$\tilde{N}=6$)
\begin{subequations}
\begin{eqnarray}
{\bar\phi}_0=-{\bar\phi}_3=b, \label{eq:fourteena}\\
{\bar\phi}_1=-{\bar\phi}_2=-{\bar\phi}_4={\bar\phi}_5=\frac{b}{1-\omega}, \label{eq:fourteenb}\\
{\bar\phi}_{n+\tilde{N}}={\bar\phi}_n, \label{eq:fourteenc} 
\end{eqnarray}
where
\begin{equation}
b=\frac{\gamma~(1-\lambda^3)}{\gamma~(1-\lambda^3)-\frac{1-\lambda^2}{\lambda}(1+\lambda^3)}, \label{eq:fourteend} 
\end{equation}
and where 
\begin{equation}
\gamma>~\gamma_1(\lambda)|_{N=3,\delta=\pi}~=\frac{1+\lambda+\lambda^3 +\lambda^4}{\lambda+\lambda^3}. \label{eq:fourteene}
\end{equation}
\end{subequations}
Having obtained the commensurate and incommensurate standing waves in our
 piecewise  smooth DNLS-like model, we now turn to their linear stability 
analysis. 
\section{Linear Stability Analysis} 
\subsection{The transfer matrix}
We consider a perturbation $u_n$ over the breather profile ${\bar\phi}_n$ of 
(\ref{eq:twoa}), (\ref{eq:ten}), (\ref{eq:eleven}) in the rotating frame and write (we drop the bar over $\phi_n$)
\begin{equation}
\psi_n=(\phi_n+u_n(t)) e^{-i \omega t}. \label{eq:fifteen}
\end{equation}
Using this in \eqref{eq:onea}, \eqref{eq:oneb} and linearising, one obtains
\begin{equation}
i \dot{u}_n + \omega u_n + u_{n+1} + u_{n-1} = 
-\gamma \{u_n - \frac{\phi_n}{2|\phi_n|}(\frac{u_n}{\phi_n}-\frac{u_n^*}
{\phi_n^*})\}\ominus(|\phi_n|-1). \label{eq:sixteen}
\end{equation}
This actually represents a linear Hamiltonian system, and we seek a harmonic 
solution of the form 
\begin{equation}
u_n(t)=a_n e^{i p t}+ b_n e^{-i p^* t},  \label{eq:seventeen}
\end{equation}
where $p$ (in general complex) is an eigenfrequency with eigenmodes specified 
by $\{a_n\},~\{b_n\}$. Substituting in \eqref{eq:sixteen}, we obtain the equations satisfied by the 
$a_n's,$ and $b_n's$ : 
\begin{subequations}
\begin{equation}
p a_n = \omega a_n + a_{n+1} + a_{n-1} + \gamma \{ a_n -
\frac{\phi_n}{2 |\phi_n|}(\frac{a_n}{\phi_n}-\frac{b_n^*}{\phi_n^*})\} \ominus
(|\phi_n|-1), \label{eq:eighteena}
\end{equation}
\begin{equation}
p b_n^* = - \omega {b_n^*} -b_{n+1}^* - b_{n-1}^*-\gamma\{b_n^*
-\frac{\phi_n^*}{2|\phi_n|}(\frac{b_n^*}{\phi_n^*}-\frac{a_n}{{\phi}_n})\}
\ominus {(|\phi_n|-1)}. \label{eq:eighteenb}
\end{equation}
\end{subequations}
\vskip .5cm
\noindent
These equations define a linear mapping
\begin{equation} 
\left(  \begin{array}{l}
 a_n \\ a_{n-1} \\ b_n^* \\ b_{n-1}^* \end{array} \right) 
\rightarrow \left(  \begin{array}{l}
 a_{n+1} \\ a_n \\ b_{n+1}^*\\ b_n^* 
\end{array}  \right) 
= A_n \left(   \begin{array}{l} 
  a_n \\ a_{n-1} \\b_n^* \\b_{n-1}^*  
\end{array}   \right), \label{eq:nineteen}
\end{equation}
where, using \eqref{eq:ten}, the matrix $A_n$ is obtained as 
\begin{subequations} 
\begin{equation}
A_n\equiv A_1^{(k)}(p)= \left(  \begin{array}{cccc}
p-\omega-\gamma~(1-\frac{1}{2b}) & -1 & -\frac{\gamma}{2b} e^{2ik\delta} & 0 \\
               1                 &  0 &                 0                & 0 \\
-\frac{\gamma}{2b}e^{-2ik\delta} &  0 & -p-\omega-\gamma(1-\frac{1}{2b}) & -1\\
                0                &  0 &               1                  & 0
\end{array}   \right), \label{eq:twentya}
\end{equation}
 
\noindent for $n=kN~~(k=0,\pm1,\pm2,\ldots)$,  and

\begin{equation}
 A_n\equiv A_2(p)=\left(  \begin{array}{cccc}
 p-\omega & -1 & 0 & 0 \\
   1      &  0 & 0 & 0  \\
   0      &  0 & -p-\omega & -1 \\
   0      &  0 &     1     &  0  
\end{array}     \right), \label{eq:twentyb}
\end{equation}
\end{subequations}
for $n \ne kN.$ 
\vskip .5cm
\noindent
In other words, the mapping from {\it high} site to its next ({\it low}) site
differs from the mapping from a {\it low} site to the next site, which may 
be a high or a low one. Since there are $(N-1)$ low sites in between two successive high sites, we obtain the {\it transfer matrix} from a {\it high} site at 
$n=kN$ to the next high site $n=(k+1)N$ as
\begin{equation}
T_k(p)={(A_2(p))}^{N-1}~ {A_1^{(k)}}(p). \label{eq:twentyone}
\end{equation}
Thus, for instance, the mapping from the column vector
\begin{equation}
X_n \equiv 
\left( \begin{array}{l}
 a_n \\ a_{n-1} \\ b_n^* \\ b_{n-1}^* 
 \end{array} \right), \nonumber
\end{equation}
based at sites $n=kN$ to the vector based at $n=(k+r)N$ is
\begin{equation}
X_{(k+r)N} = {T_{k+r-1}}(p)....{T_{k+1}}(p)~ {T_k}(p)~ X_{kN}. \label{eq:twentytwo} 
\end{equation}
\vskip .5cm
\noindent
The criterion for the determination of the spectrum of eigenfrequencies can now 
be stated as follows : a complex number $p$ is an eigenfrequency for the 
system (\ref{eq:sixteen}) if there exists a vector $X_0$ such that all the vectors 
$X_{rN}~(r=0,~\pm 1,~ \pm 2, \ldots)$ obtained from $X_0$ by successive 
applications of ${T_k}(p)$ and ${{T_{k}}^{-1}}(p)$ (with appropriate $k$) remain finite for 
$r~ \rightarrow~ \pm~ \infty$. Notice that for an {\it incommensurate} SW
solution, for which $\delta$  is not a rational multiple of $2\pi$, the 
sequence of matrices ${T_k}(p)~ (k=0,~ \pm 1,~ \pm 2,\ldots)$ never repeats itself, 
while for a commensurate solution with $\delta = ({\frac{r}{q}}) 2\pi$ one
has  (refer to (\ref{eq:twentya}) which involves $2 \delta $ rather than $\delta$),
\begin{equation}
{T_{k+s}}(p)={T_{k}}(p), \label{eq:twentythree}
\end{equation}
where $s=q$ or $\frac{q}{2}$ according as $q$ is odd or even 
($s=1$ for $\delta=0$).
In the latter situation the problem reduces completely to that of diagonalising the $4\times4$ 
{\it transfer matrix} 
\begin{equation}
R(p) \equiv {T_{s-1}}(p)\ldots{T_{1}}(p)~ {T_{0}}(p). \label{eq:twentyfour}
\end{equation}
In particular, for $\delta = 0$ or $\pi$ one has to diagonalise 
\begin{subequations}
\begin{equation}
R(p) \equiv {(A_2(p))}^{N-1}~ {A_1}(p), \label{eq:twentyfivea}
\end{equation}
where
\begin{equation}
A_1(p)= \left( \begin{array}{cccc}
p-\omega-\gamma(1-\frac{1}{2b}) & -1 & -\frac{\gamma}{2b} & 0 \\
        1                  &  0 &         0          & 0  \\
   -\frac{\gamma}{2b}      &  0 &  -p-\omega-\gamma(1-\frac{1}{2b}) & -1 \\
         0                 &  0 &          1                        &  0
\end{array}   \right), \label{eq:twentyfiveb}
\end{equation}
\vskip .5cm
\noindent
and ${A_2}(p)$ is given by \eqref{eq:twentyb}. It is easy to check from (\ref{eq:twentyb}) 
that   
\begin{equation}
{A_2(p)}^{N-1}=\left(    \begin{array}{cccc}
\frac{{\mu_1}^N-{\mu_1}^{-N}}{\mu_1-{\mu_1}^{-1}} & -\frac{{\mu_1}^{N-1}-{\mu_1}^{-N+1}}{\mu_1-{\mu_1}^{-1}} & 0 & 0  \\ \\
\frac{{\mu_1}^{N-1}-{\mu_1}^{-N+1}}{\mu_1-{\mu_1}^{-1}} & -\frac{{\mu_1}^{N-2}-{\mu_1}^{-N+2}}{\mu_1-{\mu_1}^{-1}} & 0 & 0 \\ \\
0 & 0 & \frac{{\mu_2}^N-{\mu_2}^{-N}}{\mu_2-{\mu_2}^{-1}} & -\frac{{\mu_2}^{N-1}-{\mu_2}^{-N+1}}{\mu_2-{\mu_2}^{-1}} \\ \\
0 & 0 & \frac{{\mu_2}^{N-1}-{\mu_2}^{-N+1}}{\mu_2-{\mu_2}^{-1}} & -\frac{{\mu_2}^{N-2}-{\mu_2}^{-N+2}}{\mu_2-{\mu_2}^{-1}}
\end{array}  \right), \label{eq:twentyfivec}
\end{equation}
\vskip .5cm
\noindent
where
\begin{equation}
 \mu_{1,2}=\frac{1}{2} [ \pm p-\omega-\sqrt{(p \mp \omega)^2-4} ]. \label{eq:twentyfived}
\end{equation}
\end{subequations}
\vskip .5cm
\noindent
Further, one observes that the eigenvalue of the transfer matrix $R(p)$  occur
in reciprocal pairs $(\lambda,~ \frac{1}{\lambda})$ since positive and negative 
directions along the lattice are equivalent (it is easy to check this 
explicitly for (\ref{eq:twentyfivea})). 
\vskip .5cm
\noindent 
Referring to (\ref{eq:seventeen}) one observes that an arbitrarily chosen 
value of $p$ (in 
general complex) will be an eigenefrequency of the system (\ref{eq:sixteen})
 if at least one pair of the eigenvalues of the transfer matrix lies on the 
unit circle in the complex plane. One can thus determine the entire eigenvalue
spectrum and, in addition, the eigenmodes. Evidently, the spectrum as also the 
eigenmodes depends parametrically on $\lambda,~\gamma$ (for given $N,~\delta$).
If, for some particular choice of $\lambda,~\gamma$, the entire spectrum of 
eigenfrequencies lies on the real axis then the corresponding SW solution 
given by (\ref{eq:ten}, \ref{eq:eleven}) is linearly stable. 
\vskip .5cm
\noindent
One notes from (\ref{eq:eighteena}),(\ref{eq:eighteenb}) that if $p$ is an 
eigenfrequency then so are $p^*,-p,-p^*$ (actually (\ref{eq:sixteen}) is a 
linear Hamiltonian system; see section 4.2).
Further, it follows from above that, for the commensurate SW's, the eigenmodes
are of the Bloch form. It is also apparent that the spectrum in this case  
consists of bands (as we see in sec. 5, the bands can be explicitly obtained in the present model). Starting
from any $\lambda,~\gamma$ for which the entire band spectrum is confined to 
the real axis, one can vary one of the parameters (or an appropriate function
of these) such that, as some critical parameter value is crossed the spectrum 
acquires complex eigenfrequencies. This corresponds to the onset of instability.
\vskip .5cm
\noindent
In the following, we apply these principles in sections 4.3, 5, to see how the 
bands, together with the band-edges and the transition points (see below) 
can be explicitly calculated and calculate the exact stability threshold for a couple
of concrete cases ($N=2,~\delta=\pi$ , and $N=3,~ \delta=\pi$). However, before
that we digress briefly to explain two important characteristics of the 
eigenfrequencies and the eigenmodes, namely the {\it Krein signatures} and
{\it Floquet multipliers}.
\subsection{Linear Hamiltonian systems with spatial periodicity : Krein 
signatures and Floquet multipliers}. \\
\noindent The evolution of the  $u_n$'s (eq. \eqref{eq:sixteen}) can be obtained from the Hamiltonian 
\begin{eqnarray}
H=\sum_n\{ (\omega+\gamma)~{|u_n|^2} + {u_{n+1}}u_n^* + {u_n}u_{n+1}^*+ ~~~~~~~~~~ \nonumber \\
+\frac{\gamma}{2|\phi_n|}(-|u_n|^2 + \frac{1}{2|\phi_n|^2}({{\phi_n}^2}{{u_n^*}^2} + {{u_n}^2}{{{\phi_n}^*}^2})) \ominus (|\phi_n|-1)\}. \label{eq:twentysix}
\end{eqnarray}
Note that, as a result, the antisymmetric product 
\begin{equation}
2H=-i\sum\{\dot{u_n} {u_n^*}-\dot{u_n^*} u_n \}, \label{eq:twentyseven}
\end{equation}
is conserved. Assuming a harmonic time dependence 
\begin{equation}
u_n={a_n}~ e^{i p t} + {b_n}~ e^{-i p^* t}, \label{eq:twentyeight}
\end{equation}
we find 
\begin{equation}
-i \sum(\dot{u_n}{u_n^*} -{\dot{u_n^*}}{u_n})=(p+p^*)~\{\sum(|a_n|^2 - 
|b_n|^2)\}~ e^{i(p-p^*) t}. \label{eq:twentynine}
\end{equation}
\vskip .5cm
\noindent
Thus, in order that the left hand side be conserved we must have 
\begin{equation}
\sum_n{(|a_n|^2 - |b_n|^2)}=0, \label{eq:thirty}
\end{equation}
if $p \neq p^*.$ On the other hand, for a real $p$, the signature of the 
quantity $\sum(|a_n|^2-|b_n|^2)$ which is proportional to the energy evaluated
for the solution \eqref{eq:twentyeight}, is termed the {\it Krein Signature} (see~\cite{ref3}) for frequency $p$
corresponding to the eigenmode specified by  $\{a_n\},~\{b_n\}.$ Denoting the
eigenmode by the symbol $\chi$ we express the Krein signature as $K(p;\chi).$
We have seen above that $K(p;\chi)$ is zero if $Im(p) \ne 0.$ If the 
eigenfrequency $p$ is degenerate with eigenmodes $\chi_1, \ldots, \chi_s$ then 
for each eigenvector we have a Krein signature $K(p;\chi_i)~ (i=1,\ldots s).$
More generally, we can talk of the Krein signature for a subspace, say, $V$,
spanned by a given subset of the set of eigenvectors $\chi_1,....\chi_s$, and 
call it $K(p;V)$. It is defined to be `$+$',  `$-$', or $0$ if the reduced energy 
$\frac{H}{\frac{1}{2}(p+p^*)}$ restricted to the subspace $V$ is positive 
definite, negative definite, or indefinite~\cite{ref7}. Finally, if $V$ is taken to be the 
span of all the eigenvectors $\chi_{1},\ldots,\chi_{s}$, then $K(p;V) $ is termed 
the Krein signature for the eigenfrequency $p$ and denoted simply as $K(p).$
\vskip .5cm
\noindent
The relevance of the Krein signature lies in the fact that it characterises 
critical changes in the structure of the spectrum of eigenfrequencies for the
given quadratic Hamiltonian $H$ as a function of parameters through which 
$H$ can be made to change continuously. In particular, it indicates the 
onset of bifurcations wherein the the spectrum acquires eigenfrequencies 
with non-zero imaginary parts. For instance, we have the important result~\cite{ref6,ref13} that
if $ \nu$ be a relevant parameter in $H$ such that two real isolated 
eigenfrequencies ${p_{1}}(\nu)$, ${p_{2}}(\nu)$ with opposite Krein signatures
collide at $\bar{p}$ as $\nu$ approaches a value $\bar{\nu}$, say, from below,
then after the collision as $\delta=(\nu-\bar{\nu})$ becomes positive the two 
eigenfrequencies spread out in the complex plane as $\bar{p}+{\epsilon_{1}}(\delta)
 \pm i {\epsilon_{2}}(\delta)$, where $\epsilon_{1}$ and ${\epsilon_{2}}$ are functions 
tending to zero as $\delta \rightarrow 0$ ( the converse of this statement is
also true, see~\cite{ref7}). Such collisions are responsible for the so called 
Hamiltonian Hopf bifurcation in non-linear  Hamiltonian systems - a subject
widely discussed in the literature (see~\cite{ref6,ref8,ref9,ref10} and references therein). 
\vskip .5cm
\noindent
While the eigenfrequencies and the eigenmodes of a linear Hamiltonian system
are characterised by their Krein signatures, much more can be said about the 
eigenmodes if the system happens to possess spatial periodicity. Thus, in 
\eqref{eq:twentysix}, the coefficients of the linear system are spatially periodic because
of the periodicity of the breather profile function ($\phi_n$) for the 
standing wave solution under consideration, and hence the eigenmodes have a 
Bloch-like structure, characterised by appropriate Bloch wave numbers. Thus,
the coefficients $a_n$ (as also $b_n$) in \eqref{eq:twentyeight} are of the form 
\begin{equation}
a_n = {\tau}^n A_n, \label{eq:thirtyone}
\end{equation}
where $A_n=A_{n+\bar{N}}$ ($\bar{N}$ is the spatial period ; $\bar{N}$ may be
different from $\tilde{N}$, the spatial period of the commensurate SW; 
e.g., for $\delta=\pi,~ \tilde{N}=2N$, but $\bar{N}=N.$)
, and $\tau$ 
is the Bloch multiplier. One can simplify by looking at the values of $a_n$
at $n=k~\bar{N}$ $(k=0,\pm 1, \pm 2,\ldots)$ so that, writing $a_{k \bar{N}}=
{\tilde{a}_k}$ we have 
\begin{equation}
{\tilde{a}}_k=A \lambda^k, \label{eq:thirtytwo}
\end{equation}
where $A$ is a constant and $\lambda={\tau}^{\bar{N}}$ is the {\it Floquet 
multiplier} characterising the eigenmode. As a consequence of the fact that both
the two spatial directions in the lattice are equivalent, Floquet multipliers 
necessarily occur in pairs like $(\lambda,\frac{1}{\lambda})$. Choosing an 
arbitrarily specified frequency $p$ (in general complex), the solution for 
$a_n$ (also for $b_n$) can be written as a linear combination of special 
solutions of the form \eqref{eq:thirtyone}, \eqref{eq:thirtytwo} with Floquet multipliers, say, 
$(\lambda_i,\frac{1}{\lambda_i})$ $(i=1,2,\ldots)$. If at least one of these 
pairs lies on the unit circle ($i.e.,|\lambda_i|=1$ for some $i$) then the 
chosen value of $p$ qualifies as an eigenfrequency, with these specified
${\lambda_{i}}'s$  determining the eigenmodes $\chi_j$ ($j=1,2,\ldots,s$, say)
belonging to $p$. If there be $r$ pairs of multipliers lying on the unit 
circle, then the maximum possible number of eigenmodes is $s=2r.$ However, 
$s$ can be less than $2r$ depending on degeneracies among the ${\lambda}'s.$ 
\vskip .5cm
\noindent
When a relevant parameter $\nu$ characterising the Hamiltonian is made to vary,
the eigenfrequency $p$, in general,  moves continuously in the complex plane 
and, at the same time, the Floquet multipliers also move in the complex $\lambda$-
plane. In the process, certain critical values of $\nu$ may arise when pairs 
of eigenvalues either leave or enter into the unit circle in the $\lambda$-plane. The range of values covered by $p$ for which at least one $\lambda$-pair resides on the unit circle constitutes a {\it band} while inside the band,
one may have {\it transition points} corresponding to an increase or decrease in the
numbers of such pairs as mentioned above. The transition points divide a band 
into segments and at certain special values of the parameter $\nu$, a segment 
may shrink to zero width through a collision of a pair of transition points.
As we shall see below, it is precisely  at these special values that the 
Krein signature undergoes a change indicating, possibly, the onset of an 
instability. 
\vskip .5cm
\noindent
In summary, the Krein signatures associated with the eigenfrequencies forming
a band in an infinite dimensional linear Hamiltonian system with Bloch type
eigenmodes are intimately connected with the Floquet multipliers characterising
these eigenmodes. In particular, collisions of transition points within bands
(as also collisions between band-edges) are associated with changes in the
Krein signature signifying a possible onset of instability.

\subsection{Determination of the band-edges and transition points.}
Referring to sections 4.1, and 4.2, we observe that the eigenvalues of the 
transfer matrix $R$ for a commensurate $SW$ solution, are precisely the 
Floquet multipliers for the problem. There are two pairs of these multipliers 
for any (complex) $p$, of which at least one pair has to be on the unit circle 
for $p$ to qualify as an eigenfrequency. The multipliers, in general are of the
form $(\lambda_1,~\frac{1}{\lambda_1},~\lambda_2,~\frac{1}{\lambda_2})$. 
However, for real $p$, the transfer matrix $R(p)$ is real, and a multiplier 
$\lambda$ implies others of the form $\frac{1}{\lambda},~\lambda^*,
~\frac{1}{\lambda^*}.$ Assuming that, for any given real $p$ all four 
multipliers lie on the unit circle (fig. 1, A1) one observes that, as the 
eigenfrequency is made to vary along the real line within a band, the two pairs 
move on the unit circle till one pair of multipliers collide at $+1$ or $-1$
(fig.1, A2, A3), after which that pair moves off the unit circle along the
real line, (fig.1, B1, B2) while another pair of multipliers continues to lie
on the unit circle. The degenerate configurations $A2$, $A3$ of fig.1 thus
mark {\it transition points} in the band of eigenfrequencies. At these 
transition points one has, respectively, either
\begin{subequations}
\begin{equation}
    P+2-2T=0, \label{eq:thirtythreea}
\end{equation}
or
\begin{equation}
  P+2+2T=0, \label{eq:thirtythreeb}
\end{equation}
\end{subequations}
where $P(p)$ and $T(p)$ are respectively given by 
\begin{subequations}
\begin{equation}
T(p)=\sum{\lambda_i}, \label{thirtyfoura}
\end{equation}
\begin{equation}
P(p)=\sum_{i<j}{\lambda_i \lambda_j}, \label{eq:thirtyfourb}
\end{equation}
\end{subequations}
\vskip .5cm
\noindent
in terms of the eigenvalues $\lambda_i~(i=1,2,3,4)$ of $R(p).$ Away from a 
transition point, a possible scenario consists of the pair of multipliers 
remaining on the unit circle moving along the latter till these, in turn, 
become degenerate at $+1$ or $-1$ (fig. 1, C1, C2, C3, C4) and then cease 
to lie on the unit circle (fig.1., D1, D2, D3). These correspond to 
{\it band-edges} $i.e.,$ points on the real $p$-axis where the band terminates. At these band-edges
one again has either (\ref{eq:thirtythreea}) or (\ref{eq:thirtythreeb}), but now 
lying in a different region in the $P$-$T$ plane (we do not enter here into a 
characterisation of these regions). A band on the real $p$-axis can also 
terminate as both pairs of multipliers on the unit circle collide to form a 
degenerate Krein configuration (fig.1, E1), and then all four leave the unit 
circle (fig. 1, E2). Such a band-edge is characterised by the relation 
\begin{equation}
4(P-2)=T^2,~~~|T|<4, \label{eq:thirtyfive}
\end{equation}
\noindent
between $T(p)$ and $P(p)$ defined above.
\vskip .5cm
\noindent
Having identified the conditions (in the following we refer to 
Eqns.(\ref{eq:thirtythreea}),(\ref{eq:thirtythreeb}), (\ref{eq:thirtyfive}) as conditions I, II and III respectively) characterising the
band-edges and transition points, we can then completely determine the bands
lying on the real $p$-axis by referring to the secular equation of $R(p)$ given 
by
(\ref{eq:twentyfivea})-(\ref{eq:twentyfivec}). We carry out this programme in 
the following two subsections, where we 
find that a band may have quite a complex structure involving several 
transition points dividing the band into segments. Each segment is 
characterised by a definite disposition of multipliers given by A1, B1, or B2
of fig. 1. For an eigenfrequency $p$ within a segment with disposition A1,
there are four independent eigenmodes $\chi_1, ~\chi_2, ~\chi_3,~\chi_4$, each
of the form 
\begin{equation}
\left(   \begin{array}{l}
a_{k\bar{N}} \\ a_{(k-1)\bar{N}} \\ {b^*}_{k\bar{N}} \\{b^*}_{(k-1)\bar{N}} \end{array} \right) =
e^{i k \phi} \left(  \begin{array}{c}
         \alpha \\ \beta \\ \gamma \\ \delta 
\end{array} \right) \label{eq:thirtysix},
\end{equation}
\noindent
where $ \alpha, \beta, \gamma, \delta $ can be explicitly calculated.
One thus has the Krein signatures 
\begin{subequations}
\begin{equation}
K(p,\chi_1)=K(p,\chi_2)=\sigma_1 ~~~~~(say), \label{eq:thirtysevena}
\end{equation}
\begin{equation}
K(p,\chi_3)=K(p,\chi_4)=\sigma_2 ~~~~~(say), \label{eq:thirtysevenb}
\end{equation}
\end{subequations}
and the Krein signature $K(p)$ associated with the four-dim space made up of $\chi_1,~\chi_2,~\chi_3,~\chi_4$ is `$+$', `$-$', or $0$ according as $\sigma_1~,
\sigma_2$ are both positive, both negative or of opposite signs. For 
configurations $B1,~B2$ on the other hand, there are only two independent 
eigenvectors $\chi_1, \chi_2$ for any given $p$ and one has in this case,
\begin{equation}
K(p,\chi_1)=K(p,\chi_2)=K(p)=\sigma, ~~~~~~~(say) \label{eq:thirtyeight}
\end{equation}
where $\sigma$ can be either `$+$' or `$-$'. 
\begin{figure*}
	\centering
		\includegraphics[width=0.9\columnwidth]{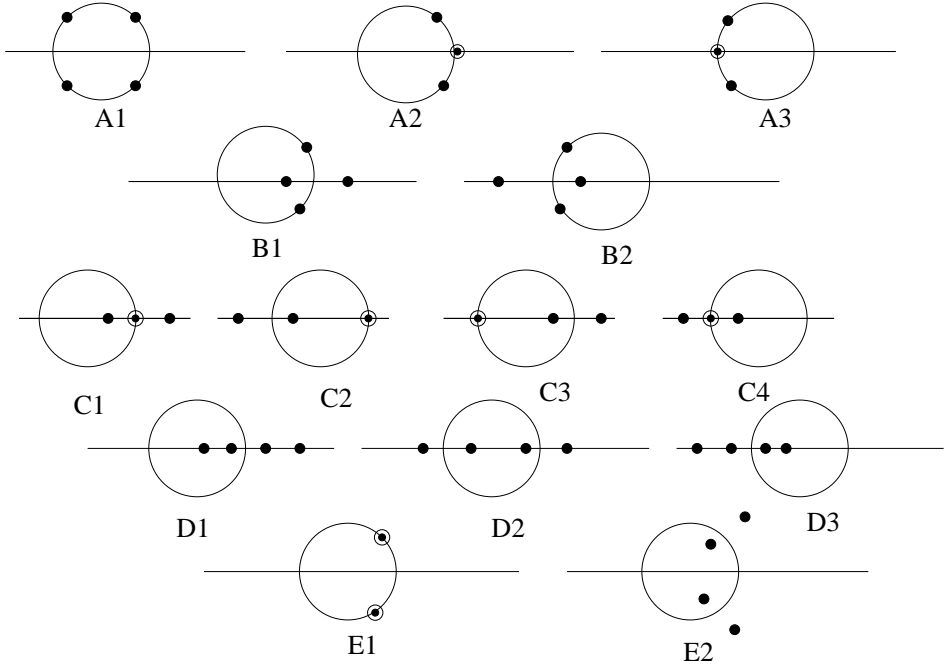}
	\caption{\label{cap:fig31}Schematic diagram depicting a number of possible dispositions of Floquet multipliers with reference to the unit circle, for a real eigenfrequency $p$}
\end{figure*}

\vskip .5cm
\noindent
A transition point or a band-edge, however, corresponds to a degenerate 
configuration of the multipliers and presents a special case. Considering a 
transition point, for example, with eigenfrequency $\bar{p}$ (say) with 
multipliers as in A2 of fig.1, the multipliers are $e^{\pm i \phi}$(say),
$1,~1$ . The transfer matrix $R(\bar{p})$ possesses only one eigenvector for
eigenvalue 1, while an independent generalised eigenvector can also be 
obtained. Correspondingly, one obtains an eigenmode $\chi$ of the form 
\begin{equation}
\left( \begin{array}{l}
a_{k\bar{N}} \\ a_{(k-1)\bar{N}} \\ b^*_{k\bar{N}} \\ b^*_{(k-1)\bar{N}} \end{array} \right)= \left( \begin{array}{l} 
\alpha \\ \beta \\ \gamma \\ \delta \end{array} \right), \label{eq:thirtynine}
\end{equation}
and the Krein signature of this mode is
\begin{equation} 
K(\bar{p},\chi) = sign(|\alpha|^2+|\beta|^2-|\gamma|^2-|\delta|^2). \label{eq:forty}
\end{equation}
\noindent
There exists a second eigenmode ${\chi}'$ associated with the same multiplier 
$+1$ that can be constructed using the eigenvector and generalised eigenvector 
of $R(\bar{p})$ referred to above, but it {\it diverges} linearly with site
index $n$. The Krein signature for this eigenmode is, in general, not defined.
It can, however, be defined for a $\bar{p}$  corresponding to a 
{\it degenarate} transition point (i.e., one where two transition points in a 
band coalesce). One finds that, for such a $\bar{p}$, $K(\bar{p},\chi)=0$ and
the Krein signature corresponding to the subspace spanned by $\chi,{\chi}'$
is also zero. 
\vskip .5cm
\noindent
As indicated in sec. 4.2, the coalescence of a pair of transition points 
within a band or of two band-edges may mark the onset of instability of the 
SW under consideration. The SW becomes unstable by way of the 
eigenspectrum acquiring
a complex part. Thus, there exist complex $p$-values for which at least one
pair of Floquet multipliers reside on the unit circle. One thereby has a band 
stretching out into the complex plane with non-zero imaginary part which 
terminates, once again, as any one of conditions (\ref{eq:thirtythreea}), 
(\ref{eq:thirtythreeb}) is satisfied. 
As we have seen in sec 4.2, the Krein signature for a $p$ lying within this part of the spectrum is zero. 
\vskip .5cm
\noindent
In the following section we present explicit results for SW's with $N=2,$ $\delta=\pi$
and $N=3,~\delta=\pi,$ exploring the detailed structure of the bands, and 
looking at the Krein signatures associated with the transition points that collide and lead
to instabilities of the SW solutions. We thereby obtain the 
{\it stability border} in the $\lambda$-$\gamma$ parameter plane demarcating 
stable from unstable SW solutions. 
\section{Details of band structure}
\subsection{ $N=2$, $\delta=\pi$}
\vskip .5cm
\noindent
We now apply the above considerations to determine the detailed structure of 
the bands including the band-edges and transition points and to obtain the 
criterion for the onset of instability. We consider first the simplest of the 
$SW$ solutions, namely those with $N=2, ~\delta=\pi.$ Here the {\it high} sites
alternate with {\it low} sites, and successive {\it high} sites have a phase
difference of $\pi$ ($SW$ solutions with $N=2,~ \delta=0$ are intrinsically
unstable).
\vskip .5cm
\noindent
As we have seen in section 3, such an SW solution is given by 
(\ref{eq:thirteena})-(\ref{eq:thirteene}) (recall that we confine ourselves in 
this paper to $\lambda>0$, 
$ \omega <-2$). One then has the transfer matrix 
\begin{equation}
R(p)={A_2}(p)~ {A_1}(p), \label{eq:fortyone}
\end{equation}
where $A_1(p),~A_2(p)$ are given by (\ref{eq:twentyfiveb}), (\ref{eq:twentyb}) 
respectively, and it is easy to set up the secular equation for $R(p)$, obtaining
\begin{subequations} 
\begin{equation}
T(p)=\sum{\lambda_i}=2 p^2 + (\omega^2 + \omega \gamma - 4), \label{eq:fortytwoa}
\end{equation}
\begin{equation}
P(p)=\sum{\lambda_i \lambda_j}=p^4 - (4+\omega^2)~p^2-2(\omega^2+\omega \gamma
-3) \label{eq:fortytwob}
\end{equation}
\end{subequations}
\vskip .5cm
\noindent
One can now locate the band-edges and transition points from conditions I, II
and III (eqns. (\ref{eq:thirtythreea}), (\ref{eq:thirtythreeb}), 
(\ref{eq:thirtyfive}) respectively). Using the notation 
\begin{equation}
              z=p^2, \label{eq:fortythree}
\end{equation}
One observes that $T$ and $P$ are polynomials in $z$ of degree $1$ and $2$ 
respectively, and finds, 
\begin{subequations}
\begin{equation}
(Condition~ I) ~~~z = z_{1,2}=\frac{1}{2}\{8+\omega^2 \pm \sqrt{(\omega^4 +
32 \omega^2 +16 \omega \gamma)}\}, \label{eq:fortyfoura}
\end{equation}
\begin{equation}
(Condition ~II) ~~~~~~~~~~~~~z=0,~ \omega^2(=z_3,~ say),~~~~~~~~~~~ \label{eq:fortyfourb}
\end{equation}
\begin{equation}
(Condition ~III)~~~z= z_4 = - \frac{\omega (\omega + \gamma)^2}{4 (\gamma + 2\omega)}~. \label{eq:fortyfourc}
\end{equation}
\end{subequations}
\vskip .5cm
\noindent


\noindent We mention that the solution (\ref{eq:fortyfourc}) satisfies 
$|T|<4$ (see (\ref{eq:thirtyfive})) only if
\begin{subequations}
\begin{equation}  
\tilde{\gamma}<\gamma<-3\omega, \label{eq:fortyfivea}
\end{equation}
\noindent where
\begin{equation}
\tilde\gamma=\frac{1}{\omega}[-2(\omega^2-4)-\sqrt(\omega^4+64)].\label{eq:fortyfiveb}
\end{equation}
\end{subequations}
\vskip .5cm
\noindent  Recall that the standing wave solution under consideration exists only for 
$\gamma>-\omega.$ In order to describe the stability characteristics of  the SW 
solution we look at (\ref{eq:fortyfoura})-(\ref{eq:fortyfourc}) 
for various fixed values of $\omega~(\omega<-2)$, and, for each $\omega$, 
consider possible values of $\gamma>-\omega$.
For any given $\omega$, the solutions (\ref{eq:fortyfourb}) are non-negative 
irrespective of $\gamma$ and hence correspond to real eigenfrequencies. 
The existence of the eigenfrequency $p=0$ for all
relevant $\omega,~\gamma$ corresponds to the fact
that the SW under consideration is a time-periodic solution of an autonomous 
set of differential equations, and relates to a phase shift of the periodic 
solution. On the other hand, the solutions (\ref{eq:fortyfoura}) are both real 
and positive for 
\begin{equation}
-\omega<\gamma<-2\omega-\frac{\omega^3}{16}~.\label{eq:new1}
\end{equation}
At $\gamma=\gamma_c\equiv -2\omega-\frac{\omega^3}{16}$, the two solutions coalesce and, for $\gamma>\gamma_c,$ they become complex, corresponding to complex 
solutions for $p$. Finally, one finds that in the range $-\omega<\gamma<\gamma_c,$ the solution (\ref{eq:fortyfourc}) subject to (\ref{eq:fortyfivea}), if it 
exists, is positive, leading once again to real  eigenfrequencies $p$. 
\vskip .5cm
\noindent 
These observation lead us to conclude that all the transition points and 
band-edges, and hence the entire eigenfrequency spectrum, lie on the real 
$p$-axis for any given $\omega~(<-2)$ and $\gamma$ 
satisfying \eqref{eq:new1}. Indeed, for such a
($\omega,~\gamma)$ the spectrum in the right half plane consists of one or two 
bands with transition points
in their interior. For the sake of illustration we assume that 
$\omega$ satisfies $\omega^2<8$ and, for such a given $\omega,$ look at 
different values of $\gamma$ in the range
(\ref{eq:fortyfivea}). In this case, $z_4$ is positive and satisfies (\ref{eq:thirtyfive}), and one has
\begin{subequations}
\begin{equation} 
z_3<\bar{z}(\equiv {\bar{p}}^2)<z_4,  \label{eq:fortysixa}
\end{equation}
where
\begin{equation} 
\bar{p} =\sqrt{\frac{1}{2}(\omega^2+8)}. \label{eq:fortysixb}
\end{equation}
\end{subequations}
\noindent
Considering, first, values of $\gamma$ in the range 
\begin{equation}
\tilde{\gamma}<\gamma<-2\omega-\frac{\omega^3}{16}(<-3\omega), \label{eq:fortyseven}
\end{equation}
\noindent
the band of eigenfrequencies
extends on the real line from $p=0$ to $p=\sqrt{z_4}$ (fig. 2(a); for the sake of convenience, we consider only the right half 
complex plane of eigenfrequencies; the part of the spectrum contained in the 
left half plane is obtained by reflection about the imaginary axis).
\vskip .5cm
\noindent
The transition points in the band are at B ($p=\omega~(=\sqrt{z_3})~$), and at C, C' ($p_{1,2}\equiv\sqrt{z_{1,2}}$),
the latter two being the values of $p$ obtained from condition I. As 
$\gamma$ is increased towards the value 
$\gamma_c$ (for the chosen value of $\omega$ satisfying 
$\omega^2<8$, the condition for $z_3$ to satisfy $z_3<\bar{z}$) the transition 
points C, C' gradually come closer till, at $\gamma=\gamma_c$ they coalesce at A ($p=\bar{p}=\sqrt{\frac{1}{2}(\omega^2+8)}$) (fig. 2(b)). As $\gamma$ is made to 
cross $\gamma_c$ from below, the spectrum acquires a complex part and looks 
as in fig. 2(c).

\begin{figure*}
	\centering
		\includegraphics[width=0.9\columnwidth]{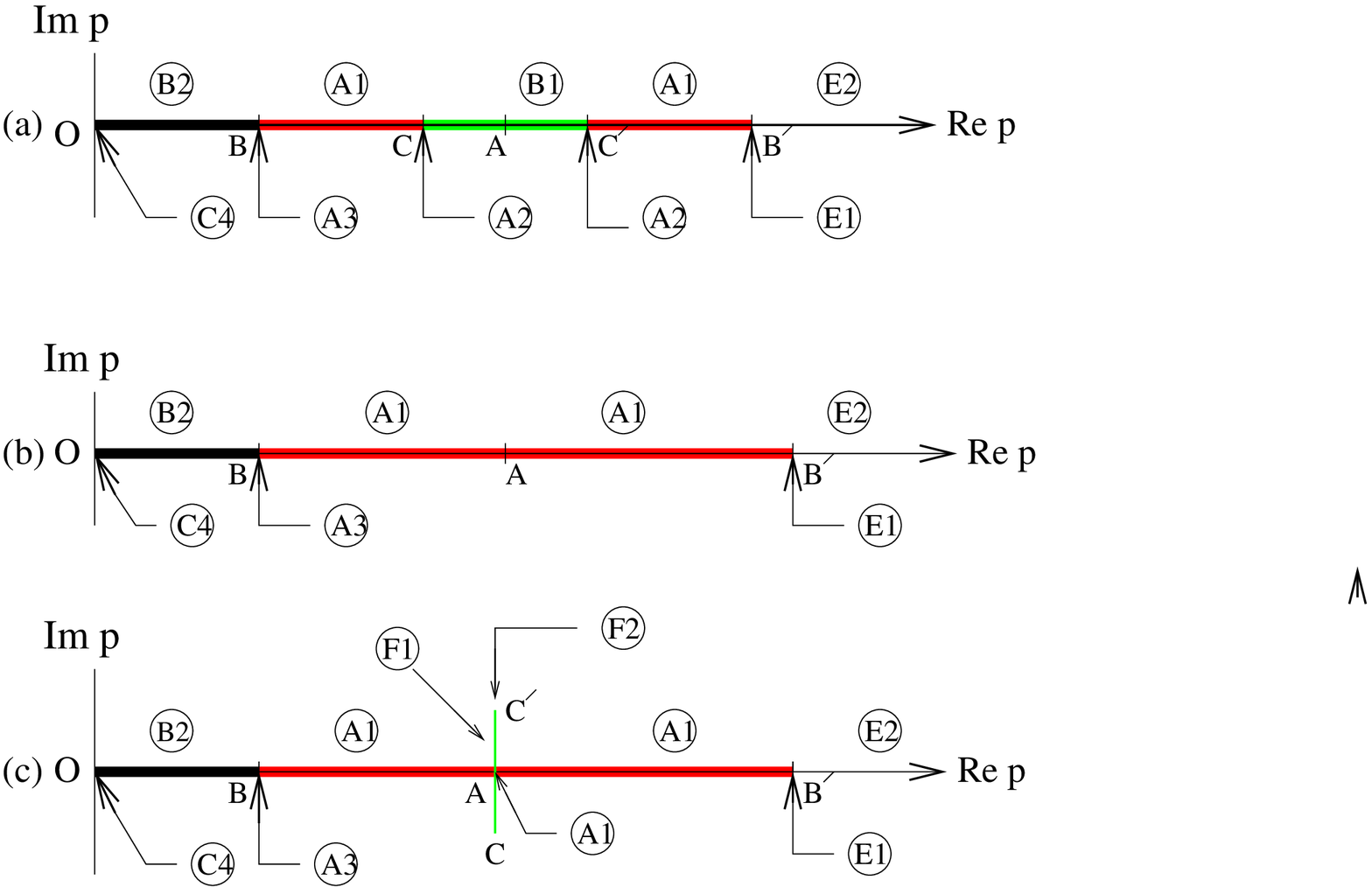}
	\caption{\label{cap:fig32}Schematic diagram of the spectrum for the standing wave solution with $N=2,~\delta=\pi$, for an appropriately chosen $\omega$ and for three values of $\gamma$, respectively less than, equal to, and greater than the critical value, see text; dispositions of the Floquet multipliers for the different band segments are indicated in accordance with notation in fig. 1}.
\end{figure*}

\vskip .5cm
\noindent
As mentioned above, Fig. 2(a,b,c), depicts 
schematically the detailed structure of the band for three different values of 
$\gamma$, namely $\gamma=\gamma_c-\epsilon,$ $\gamma=\gamma_c,$ 
$\gamma=\gamma_c+\epsilon$ respectively, where $\epsilon$ is
sufficiently small. The dispositions of the Floquet multipliers in the 
different segments of the band are indicated in each case, in accordance with 
the notation of fig. 1 and fig. 3 (see below). One observes that at the critical 
value $\gamma=\gamma_c$ (fig. 2(b)) the segment CC' of fig. 2(a) shrinks to the 
point $A$, i.e.,  the two transition points C,C' corresponding to the 
solutions of condition I (configuration A2 of fig. 1) collide and, for 
$\gamma>\gamma_c$ (fig. 2(c)), these move off the real line into the complex plane.
The spectrum thereby includes the segment CAC' stretching into the complex plane, and the standing wave 
solution becomes unstable. At the point $A$ (fig. 2(c)), the disposition of multipliers 
is like A1 of fig. 1, i.e, both pairs reside on the unit circle. For any $p$
in the interior of the segments AC or AC', the disposition of 
multipliers is like that marked F1 in fig. 3 where one pair continues to 
reside on the unit circle while the other pair moves off into the complex 
$\lambda$-plane (the stretches AC and AC' in fig. 2(c) are not actually straight, but 
curved lines). At the edges C, C' (fig. 2(c)), condition I is again satisfied (but with 
$T,~ P$ lying in a region of the $P$-$T$ plane different from the one 
corresponding to C or C' in fig. 2(a)), and the disposition of multipliers is 
like F2 in Fig. 3.

\begin{figure*}
	\centering
		\includegraphics[width=0.9\columnwidth]{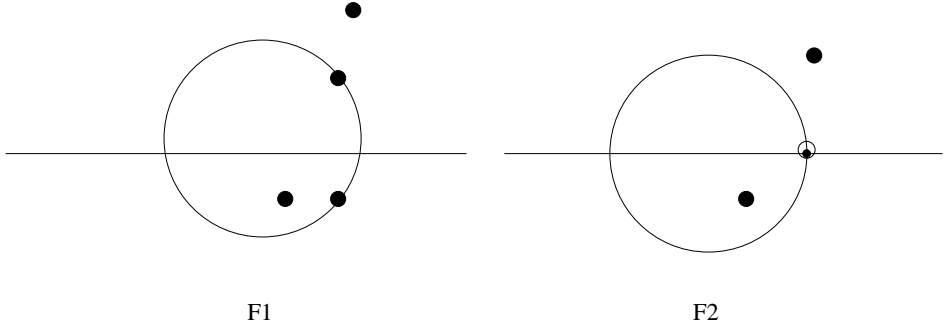}
	\caption{\label{cap:fig33}Schematic diagram of a pair of possible dispositions of Floquet multipliers for a complex $p$. }
\end{figure*}

\vskip .5cm
\noindent
Let us consider, for $\gamma=\gamma_c-\epsilon$, two points in the spectrum, 
one with $p=p_1-\delta$  (slightly left of
C) and the other with $p=p_2+\delta$ (slightly right of C')
where $\delta$ is small. For each of these points there are two pairs of multipliers on the unit circle, of which one
pair is close to $+1$. For $p=p_1-\delta$, we denote the eigenmodes of the 
pair located away from $+1$ as $\chi_1,~\chi_2$ (they are complex conjugate to 
one another), and the remaining two eigenmodes ( for multipliers close to $+1$)
as $\chi_3,~\chi_4$ (again mutually complex conjugate). Similarly, for $p=p_2+\delta$ we denote the corresponding eigenmodes as $\chi'_1$, $\chi'_2$, 
$\chi'_3$ and $\chi'_4$. All the eigenmodes have the general form
\begin{equation}
\left( \begin{array}{l}
a_{2k} \\ a_{2k-1} \\ {b^*}_{2k} \\ {b^*}_{2k-1} \end{array} \right)= e^{i k\phi}\left( \begin{array}{l}
\alpha \\ \beta \\ \gamma \\ \delta \end{array} \right), \label{eq:fortyeight}
\end{equation}
for which the Krein signature is 
\begin{equation}
\sigma=sign(|\alpha|^2+|\beta|^2-|\gamma|^2-|\delta|^2). \label{eq:fortynine}
\end{equation}
\noindent
In the present case all the eigenmodes can be explicitly obtained and one finds
\begin{subequations}
\begin{equation}
K(p_2+\delta;\chi'_1)=K(p_2+\delta;\chi'_2)=K(p_1-\delta;\chi_1)=K(p_1-\delta;\chi_2)=~- \label{eq:fiftya}
\end{equation}
\begin{equation}
K(p_2+\delta;\chi'_3)=K(p_2+\delta;\chi'_4)=-K(p_1-\delta;\chi_3)=-K(p_1-\delta;\chi_4)=~-. \label{eq:fiftyb}
\end{equation}
\end{subequations}
\noindent
Thus, the Krein signatures associated with the eigenfrequencies $p_1-\delta$ and $p_2+\delta$ are respectively
\begin{subequations}
\begin{eqnarray}
K(p_1-\delta)=0,   \label{eq:fiftyonea} \\
K(p_2+\delta)=-.      \label{eq:fiftyoneb}
\end{eqnarray}

\noindent
These are also the signatures for all $p$ in the interior of the segments 
C'B' and BC respectively. As regards the signatures for the transition 
points C and C' $(\delta=0)$ the situation is slightly different. In case 
of C ($p=p_1$), for instance, there are again two eigenmodes $\chi_1,~\chi_2$ 
(complex conjugate of each other) of the form (\ref{eq:fortyeight}). For the 
multiplier $+1,$ one eigenmode ($\chi_3$, say) is again of the form 
(\ref{eq:fortyeight}) with $(\phi=0)$, while there exits another eigenmode 
$\chi_4$ that diverges linearly with site index $n$. The signature for 
$\chi_4$ is not well defined while that for $\chi_1,~\chi_2,~\chi_3$ are as 
in (\ref{eq:fiftya},\ref{eq:fiftyb}). Similar statements apply for the 
transition point C' $(p=p_2)$. 
\vskip .5cm
\noindent
At the collision ($\gamma=\gamma_c$, fig. 2(b))
where C and
C' coalesce at A, the Krein signatures for any eigenfrequency in the 
interior of AB' and BA are again as in 
(\ref{eq:fiftyonea},\ref{eq:fiftyoneb}) respectively. As regards the 
point A $(p=\bar{p},~\gamma=\gamma_c)$ 
the signature for the two eigenmodes $\chi_1,~\chi_2$ corresponding to 
the pair of multipliers located away from $+1$ is again `$-$', while that
for $\chi_3$ is $\sigma=0$. The other eigenmode $\chi_4$ for the multiplier 
$+1$ again diverges linearly; however, now its signature is well defined and 
one finds that the reduced energy restricted to the subspace spanned by 
$ \chi_3, ~\chi_4$ is an indefinite quadratic form, 
 so that one has, for $\gamma=\gamma_c,$
\begin{equation}
K(\bar{p})=0. \label{eq:fiftyonec}
\end{equation} 
\end{subequations} 
\vskip .5cm
\noindent 
Summarising, we observe that for each of the colliding transition points C, C' there exits 
a {\it relevant} Krein signature corresponding to 
$\chi_3,~\chi'_3$ respectively which are opposite
and, at the collision, this relevant signature {\it reduces to zero}. Another way of characterising the collision is by referring to the Krein signatures for
eigenfrequencies belonging to the interiors of 
the segments BC and B'C' respectively.
As seen from (\ref{eq:fiftya}, \ref{eq:fiftyb}), these are respectively
$0$ and $'-'$. In other words, an open set of eigenfrequencies with signature 
$\sigma=0$ is embedded
within the spectrum both before and at the collision (see~\cite{ref14}).
\vskip .2cm 
\noindent
On the other side of the collision 
($\gamma>\gamma_c$, fig. 2(c)), the Krein signatures 
for eigenfrequencies in the interiors of BA, AB' remain the same (i.e., $0$ 
and $'-'$ respectively), while that for the entire complex segment CAC' one has 
$\sigma=0$.
\begin{figure*}
	\centering
		\includegraphics[width=0.9\columnwidth]{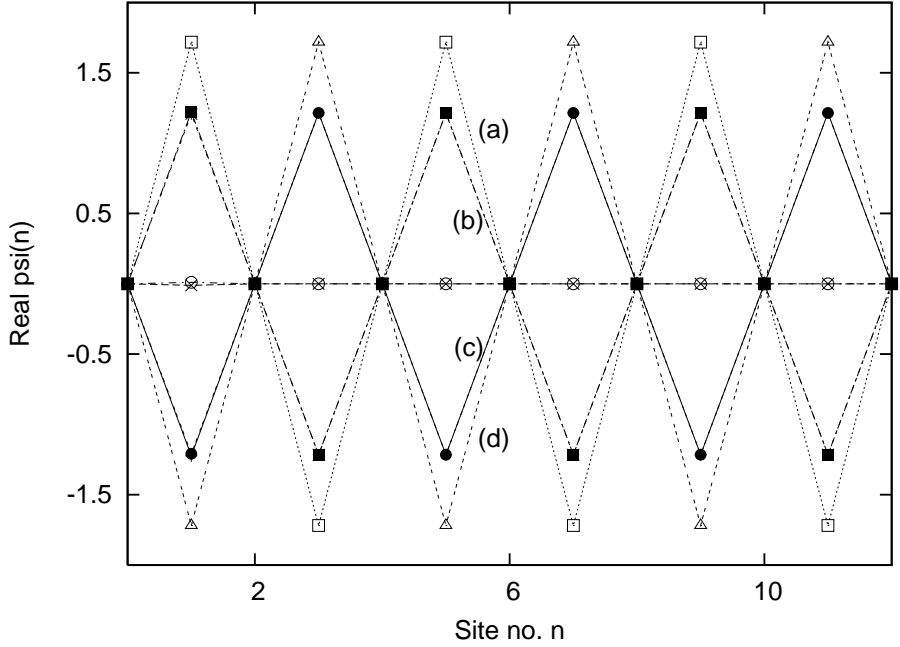}
	\caption{\label{cap:n2st}Numerical integration of \eqref{eq:onea}, \eqref{eq:oneb} to exhibit stability of $N=2,~\delta=\pi$, SW solution ($\gamma=5.9765$, slightly below the critical value) starting from the profile (as function of lattice site index $n$) at (a) $t=0$ up to $\tau=5000T$ (the new profile coincides with (a)); (b) corresponds to $\tau+\frac{T}{8}$, and $\tau+\frac{7T}{8}$, (c) to $\tau+\frac{3T}{8}$, and $\tau+\frac{5T}{8}$, and (d) to $\tau+\frac{4T}{8}$.}
\end{figure*}

\begin{figure*}
	\centering
		\includegraphics[width=0.9\columnwidth]{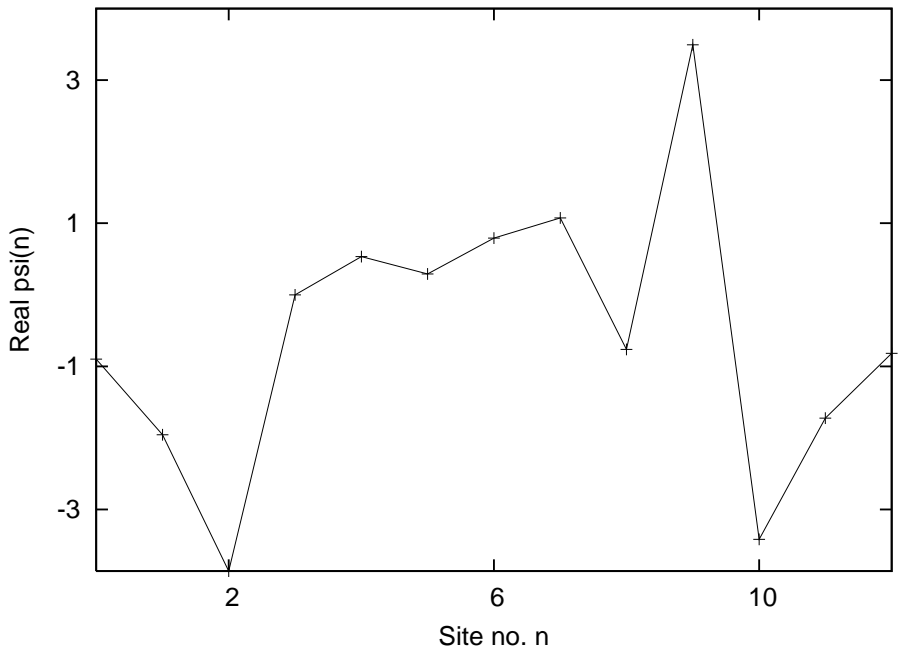}
	\caption{\label{cap:n2unst}Numerical integration as in fig. 4, but with $\gamma=5.9767$, slightly above the critical value, where the profile obtained for $\tau=1500T$ is shown; it differs markedly from the breather profile, similar to (a) in fig. 4, at $t=0$.}
\end{figure*}
\vskip .2cm 
\noindent
We thus have a complete description, both in terms of Floquet multipliers and Krein signatures, of the band structure and of the collision causing 
the onset of instability. As mentioned earlier,
for every fixed $\lambda$ (and hence 
$\omega=-\lambda-\frac{1}{\lambda})$ there exits
a critical value of $\gamma$ namely $\gamma=\gamma_c(\omega)=-2\omega-
\frac{\omega^3}{16}$
such that, as $\gamma$ crosses $\gamma_c$ from
below, the SW breather becomes unstable. We show in fig. 4, fig. 5 this loss in 
stabilty for $\lambda=0.5~(\omega=-2.5,~\gamma_c~\approx 5.9766)$ where we 
display the SW profile (eq.(\ref{eq:twoa}), (\ref{eq:thirteena})-(\ref{eq:thirteene})),
at $t=0$ as also the profile obtained 
by numerical integration of (\ref{eq:onea}), (\ref{eq:oneb}) up to a time $\tau$ 
(say, see captions) for two values of $\gamma,$ one just below and the other just above 
$\gamma_c.$ One observes in these figures that the profile remains remarkably
stable in fig. 4 for $\tau$ as large as $5000T$,
while in fig. 5, the profile quickly breaks up 
even at $\tau=1500T$ ($T=$ breather time period).
\vskip .2cm
\noindent For each fixed value of $\lambda$ $(0<\lambda<1;
~\omega<-2)$ the SW remains stable for a window of $\gamma$-values given by $\gamma_1(\lambda)<\gamma<\gamma_c(\lambda)$.
As $\lambda$ is made to increase from $0$ to $1$ 
the width of the window decreases, i.e., for 
decreasing breather frequency, the onset of instability occurs at progressively lower values of the strength of nonlinearity $\gamma$ (fig. 6).

\begin{figure*}
	\centering
		\includegraphics[width=0.9\columnwidth]{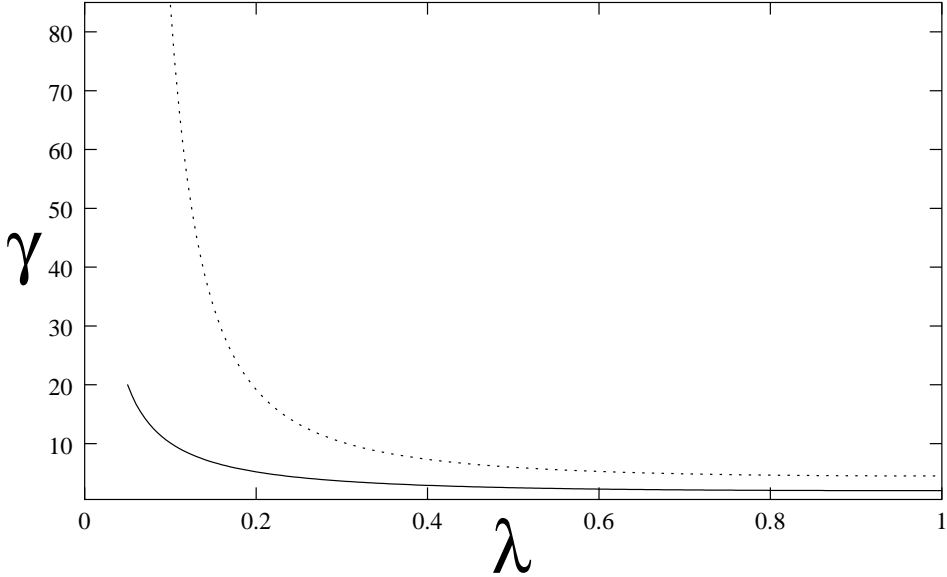}
	\caption{\label{cap:fig36}Window of stability of the $N=2, \delta=\pi$ breather; the curve with the full line corresponds to the threshold value ($\gamma_1(\lambda)$) of $\gamma$ below which the SW solution does not exist, while the dotted curve corresponds to the stability border; for any fixed $\lambda$, the vertical distance between the two curves depicts the $\gamma$-window of stability of the SW solution.}
\end{figure*}
\vskip .2cm 
\noindent

\vskip .5cm
\noindent
It is interesting to compare, for the sake of 
completeness, a different type of collision of eigenfrequencies, namely one 
where the collision
does {\it not} result in the appearance of complex
frequencies. Consider, for instance, $\omega=\omega_0=-3$ and 
$\gamma=\gamma_0=\frac{23}{3}$. With these values of $\omega$ and 
$\gamma$ (\ref{eq:fortyfoura}), (\ref{eq:fortyfourb}) give $z_1=z_3=9.$ $i.e.,
~p=\pm 3$. Fig. 7(a,b,c) shows
schematically the eigenfrequency spectrum for 
$\omega=\omega_0=-3,$ and for $\gamma=\gamma_0-\epsilon,~ \gamma=\gamma_0,~
\gamma=\gamma_0+\epsilon$ respectively, where $\epsilon$ is sufficiently 
small.

\begin{figure*}
	\centering
		\includegraphics[width=0.9\columnwidth]{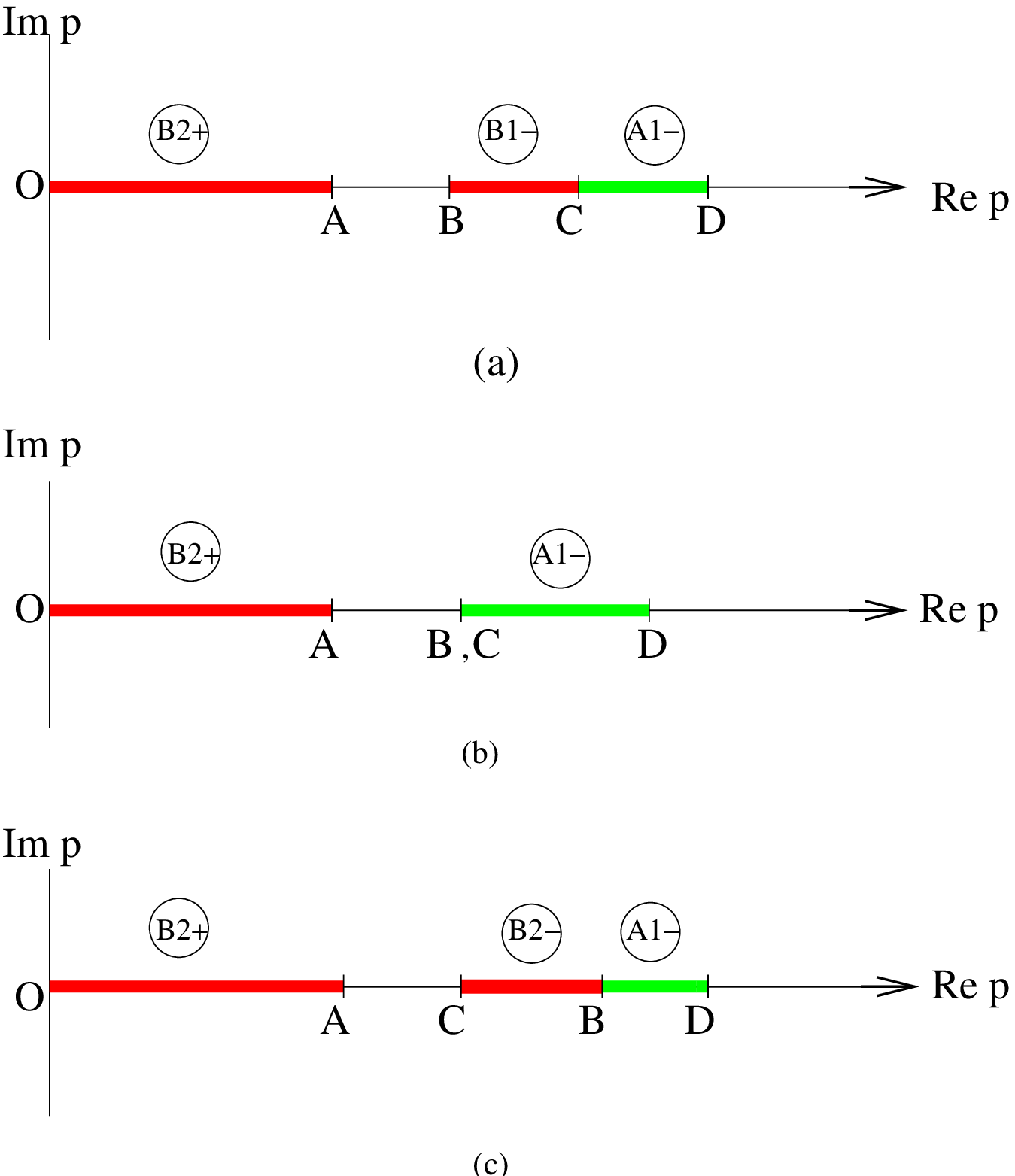}
	\caption{\label{cap:fig37}Schematic diagram of spectrum for $N=2, \delta=\pi$, SW solution similar to fig. 2, but now with $\omega=-3$, and three values of $\gamma$, respectively less than, equal to, and greater than $\frac{23}{3}$.}
\end{figure*}

\vskip .5cm
\noindent
\noindent The spectrum now consists of {\it two} bands, and the various parts of the spectrum  marked 
by transition points and band-edges have once again been marked with labels indicating the respective dispositions of Floquet multipliers in accordance with notation of fig. 1. In fig. 7(a),
for instance ($\gamma=\frac{23}{3}-\epsilon$), one band
extends from O($\omega=0$) to A ($\omega=\sqrt{z_2}$, refer to eqns.(\ref{eq:fortyfoura})-(\ref{eq:fortyfourc})), 
with only one pair of multipliers residing on the unit circle for each $p$ within this band (disposition B2 in fig.1) and the Krein signature
for such a $p$ is found to be `$+$'. We have thus labelled the band `$B2+$'. The spectrum for 
$\omega=-3$, $\gamma=\frac{23}{3}-\epsilon$ contains a second band extending from B ($p=-\omega=\sqrt{z_3}$) to D($p=\sqrt{z_4}$) through C 
($p=\sqrt{z_1}$), the latter being a transition point satisfying (\ref{eq:thirtythreea}).
For small $\epsilon$, the width of the segment BC is small and it is 
labelled `B1-' to denote that the disposition of the Floquet multipliers 
for any $p$ in the interior of the segment is B1 (fig. 1), and the Krein 
signature is `$-$'. In a similar manner, the segment CD receives the label `A1-'.
\vskip .5cm
\noindent
As $\epsilon \rightarrow 0$ i.e., $\gamma$ 
approaches the value $\frac{23}{3}$ from below (note that
this value is less than $\gamma_c=7.6875,$ the 
critical value corresponding to $\omega=-3$ at 
which one has the collision $z_1=z_2$) there occurs a collision between B 
and C, but now the two signatures involved are the {\it same} 
and there does not appear complex eigenfrequencies as a result of the 
collision, as seen in fig. 7(b,c). We observe from fig. 7(c) that
the points B, C just pass through each other, 
and now the eigenfrequencies $\sqrt{z_1},~ \sqrt{z_2}$ form the band-edges.
\vskip .5cm
\noindent
Observe that, for $\omega=-3,$ the critical collision, leading to instability occurs for 
$\gamma=\gamma_c=7.6875,$ when in fig. 7(c), the
band-edges A and C approach each other. As
seen from this figure, the Krein signatures involved are now opposite, and the spectrum acquires a complex part as $\gamma$ crosses $\gamma_c$ from below. 
\vskip .5cm
\noindent
In summary, we have obtained the detailed band structure for $N=2,~ 
\delta=\pi$, and have characterised the various bands and band segments 
in terms of the Floquet multipliers and Krein signatures, and have 
distinguished between two types of collisions in terms of the Krein signatures 
of the colliding bands or segments.

\subsection{Band Structure: N=3, $\delta=\pi$}
As another example of a collision of intraband transition points, we consider commensurate SW solutions with $N=3,$ $\delta=\pi$ (solutions with
$N=3,~\delta=0$ are again intrinsically unstable) given by (\ref{eq:fourteena})-(\ref{eq:fourteene}), where 
explicit formulae can once again be worked out. Using the notation (\ref{eq:fortythree})
, we now have:
\begin{subequations}
 \begin{equation}
T=(-5\omega+\frac{2}{1-\omega}-\gamma)z-(1-\omega)^2(\omega+2) - (\omega^2-1)\gamma, \label{eq:fiftytwoa}
\end{equation}
\begin{eqnarray}
P=-z^3+(2\omega^2+8-\frac{2(1+\gamma)}{1-\omega}){z^2}+(-\omega^4 +\frac{4\omega^3}{1-\omega}-9+ 
2\gamma\frac{\omega^2+\omega+2}{1-\omega})z+ \nonumber \\
 (2(\omega^3-3\omega+1)+2\gamma(\omega^2-1)). ~~~~~~~~~~~~ \label{eq:fiftytwob}
\end{eqnarray}
\end{subequations}
\noindent
We now have, in general, three eigenfrequencies (we once again restrict to 
$Re(p)\ge 0$) satisfying condition I (eq. (\ref{eq:thirtythreea})), and three others satisfying 
condition II (eq.(\ref{eq:thirtythreeb})), of which one is the trivial solution $p=0$, while the
number of eigenfrequencies satisfying condition III depends on $\gamma,
~\omega.$ We choose, for the sake of concreteness, $\lambda=0.5$ 
$(\omega=-2.5)$. Then two of the solutions to (\ref{eq:thirtythreea}) undergo a collision for $\gamma=\gamma_c=4.75585$ at $p=1.5750.$ We show
schematically in fig. 8(a,b,c) the detailed band
structures for three values of $\gamma,$ namely
$\gamma_c-\epsilon$, $\gamma_c$, $\gamma_c+\epsilon$ respectively, where $(\epsilon>0)$ is small. 

\begin{figure*}
	\centering
		\includegraphics[width=0.9\columnwidth]{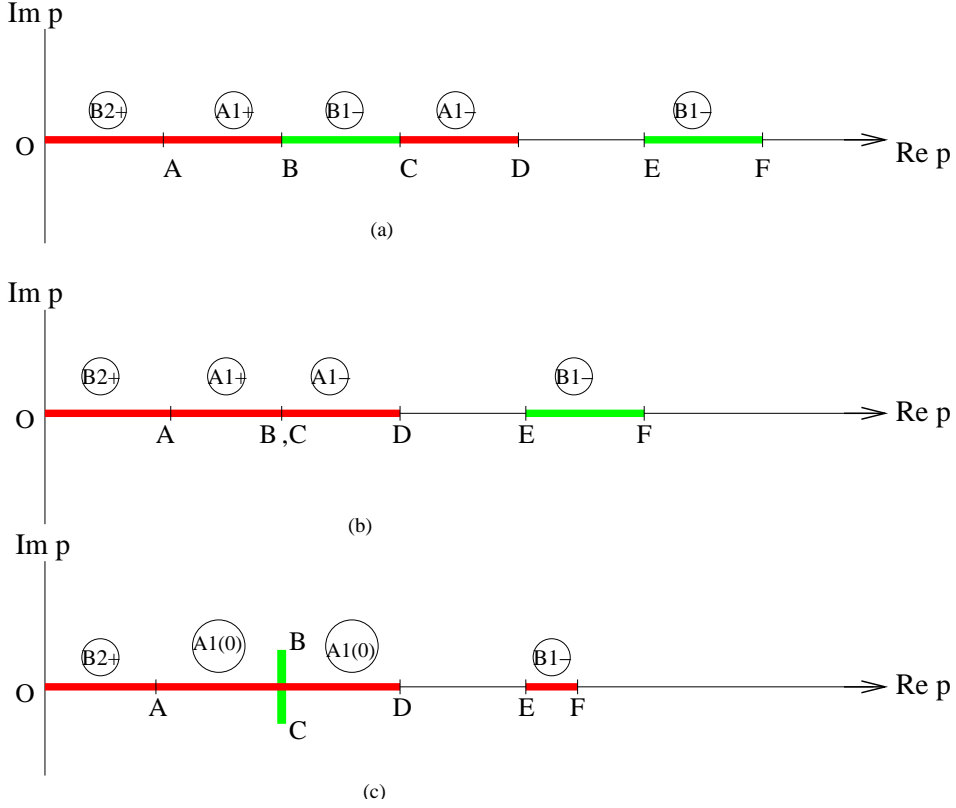}
	\caption{\label{cap:fig38}Schematic diagram, similar to fig. 2 and fig. 7, of the spectrum for the $N=3,\delta=\pi$ SW solution, with $\omega=-2.5$, and three values of $\gamma$, resp. smaller than, equal to, and larger than the critical value (see text). }
\end{figure*}

For each of these three values of $\gamma,$ we have two bands separated
by a band gap, together with a number of transition points within one of the bands. As in
fig. 7 we mark each of the band segments with a label indicating the 
disposition of Floquet multipliers  as also the Krein signature characterising 
that segment. In fig. 8(a,b,c) the
points B, C, E correspond to condition I, while O, A, F correspond to condition II, and D corresponds to condition III. The stretch from 
 D to E is a band gap. One observes that, at
$\gamma=\gamma_c,$ the transition points B and 
C within the first band collide and then stretch out into the complex plane (the 
segment from B to C in the complex plane in fig. 8(c) is not 
actually straight but is slightly curved). The label in the segment from A to D in fig.  8(c) is meant to imply that the Krein signature
for any $p$ within this segment is $0$ (i.e., the reduced energy is an indefinite quadratic form on the space spanned by the eigenmodes) while the 
disposition 
of the Floquet multipliers is similar to A1 in fig. 1. We note from fig. 8(b) 
that the two colliding segments bear opposite Krein signatures, as a result of 
which they stretch out into the complex plane on the other side of the 
collision.

\noindent
Figures 9, 10 show the results of numerical integration of an initial standing wave 
profile with $\lambda=0.5$, and with $\gamma=$4.75580$(<\gamma_c)$ and  
$\gamma$=4.75590$(>\gamma_c)$ respectively, where we find that the
SW solution indeed gets destabilised as $\gamma$
crosses $\gamma_c$ from below.

\begin{figure*}
	\centering
		\includegraphics[width=0.9\columnwidth]{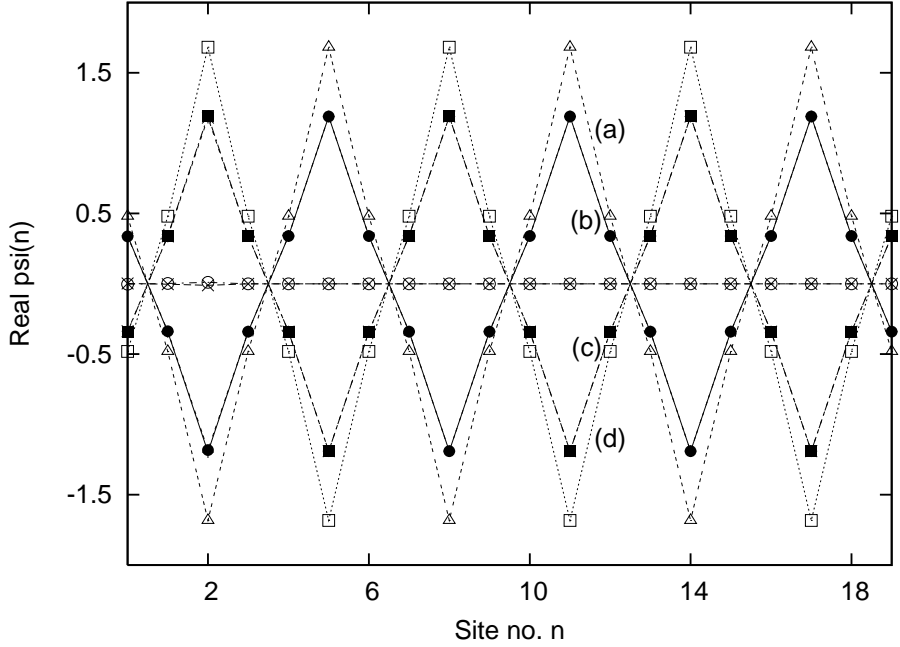}
	\caption{\label{cap:n3st}Numerical integration of SW solution for $N=3,~\delta=\pi$, with $\lambda=.5$, $\gamma=4.75580$ (slightly less than $\gamma_c$); for explanation, see caption to fig. 4.}
\end{figure*}

\begin{figure*}
	\centering
		\includegraphics[width=0.9\columnwidth]{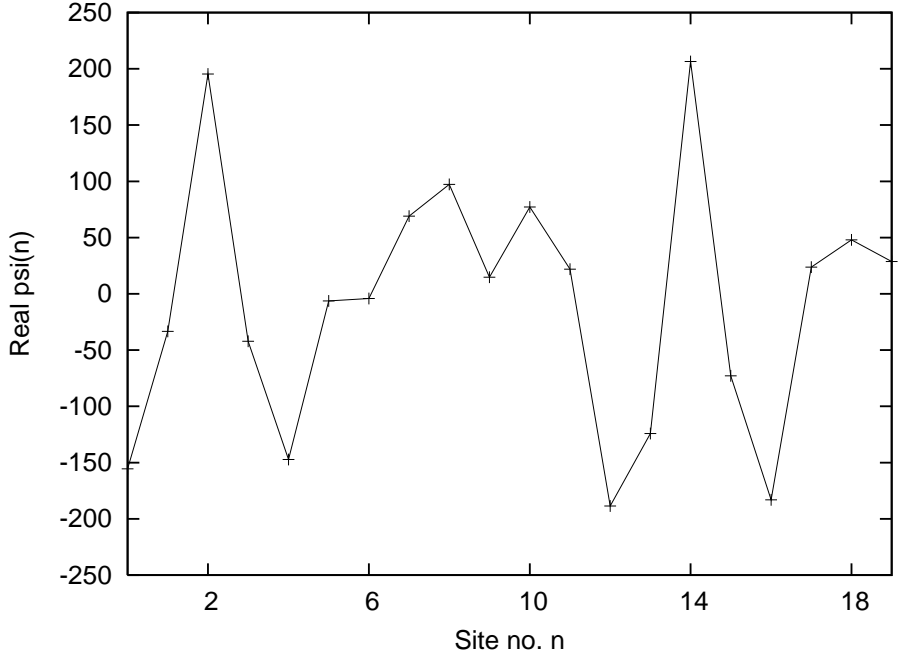}
	\caption{\label{cap:n3unst}Numerical integration of SW solution for $N=3,~\delta=\pi$, with $\lambda=.5$, $\gamma=4.75590$ (slightly greater than $\gamma_c$); for explanation, see caption to fig. 5.}
\end{figure*}
\noindent
Fig. 11 depicts the region of stability of the standing wave solution under consideration in the
$\gamma-\lambda$ parameter plane, where we once again find that, for 
decreasing breather frequency, the SW gets destabilised at progressively lower 
values of the non-linearity parameter $\gamma.$

\begin{figure*}
	\centering
		\includegraphics[width=0.9\columnwidth]{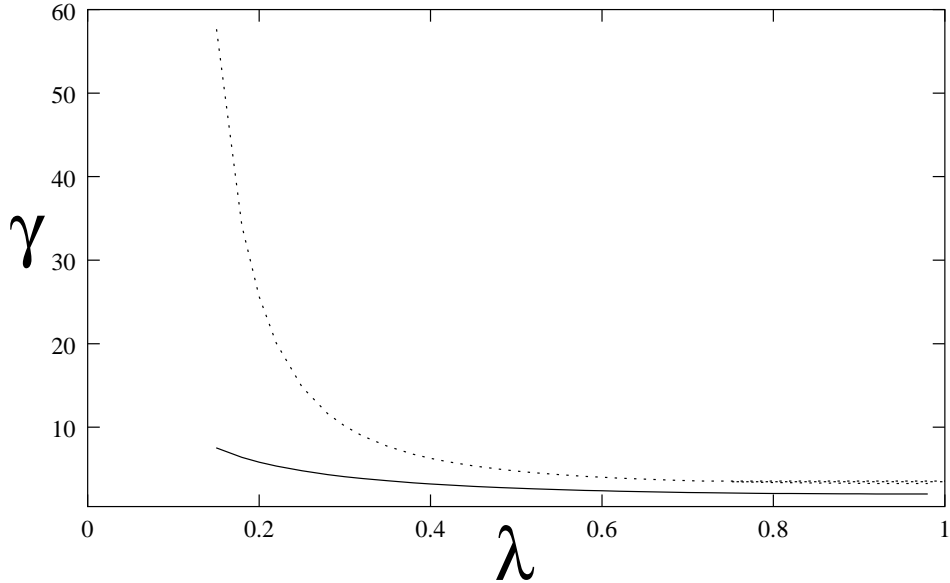}
	\caption{\label{cap:fig311} Stability window for the $N=3, \delta=\pi$, SW solution; for explanation see caption to fig. 6}
\end{figure*}
  
\vskip .5cm
\noindent In conclusion, we observe that, away from the linear and the anti-continuous limits, there exist dynamically stable commensurate standing wave solutions in our DNLS-like model, in contrast to DNLS breathers close to the above two limits. The frequency spectrum and eigenmodes of the linearised systems around these SW solutions present interesting features that can be described in terms of the dispositions of  Floquet multipliers as also the Krein signatures associated with the various segments of the band spectrum. The onset of instability occurs not necessarily through collisions of band-edges, but also through collisions of intra-band transition points.
\vskip .5cm
\noindent In a future communication, we shall study exact commensurate and incommensurate SW solutions in a piecewise linear NDKG model and look into their stability problem, with special reference to incommensurate SW solutions. The possibility of multibreaher solutions involving an infinite array of breathers with random spatial distribution will also be investigated in these exactly solvable models.

\end{document}